\documentclass[twocolumn,superscriptaddress,floatfix,showpacs,aps,longbibliography,prb,10pt]{revtex4-1}
\usepackage[english]{babel}
\usepackage{graphicx}
\usepackage{amsmath}
\usepackage{amssymb}
\usepackage{bm}
\usepackage{color}

\DeclareMathOperator{\e}{e}

\newcommand{\ket}[1]{|#1\rangle}

\usepackage{color}

\begin{document}

\title{Effects of disorder on Coulomb-assisted braiding of Majorana zero modes}
\author{I. C. Fulga}
\affiliation{Instituut-Lorentz, Universiteit Leiden, P.O. Box 9506, 2300 RA Leiden, The Netherlands}
\author{B. van Heck}
\affiliation{Instituut-Lorentz, Universiteit Leiden, P.O. Box 9506, 2300 RA Leiden, The Netherlands}
\author{M. Burrello}
\affiliation{Instituut-Lorentz, Universiteit Leiden, P.O. Box 9506, 2300 RA Leiden, The Netherlands}
\author{T. Hyart}
\affiliation{Instituut-Lorentz, Universiteit Leiden, P.O. Box 9506, 2300 RA Leiden, The Netherlands}
\date{August 2013}

\begin{abstract} 
Majorana zero modes in networks of one-dimensional topological superconductors obey non-Abelian braiding statistics. Braiding manipulations can be realized by controlling Coulomb couplings in hybrid Majorana-transmon devices. However, strong disorder may induce accidental Majorana modes, which are expected to have detrimental effects on braiding statistics. Nevertheless, we show that the Coulomb-assisted braiding protocol is efficiently realized also in the presence of accidental modes. The errors occurring during the braiding cycle are small if the couplings of the computational Majorana modes to the accidental ones are much weaker than the maximum Coulomb coupling. 
\end{abstract}

\maketitle

\section{Introduction}

Majorana zero modes appear at domain walls between the topologically distinct phases that characterize one-dimensional superconductors.\cite{Kit01} The search for these quasiparticles is motivated by their non-Abelian statistics \cite{Moo91,Read2000,Iva01, Ali11, Ste04} and the perspective they offer in quantum computation.\cite{Nay08, Hyart13} The topologically nontrivial  phase  can be realized with the help of an effective $p$-wave pairing in a spin-orbit coupled nanowire, proximity coupled to a superconductor,\cite{Lut10,Oreg10} and first signatures of Majorana modes have been reported in these setups.\cite{Mou12, Das12} Other systems supporting Majorana modes include the edge of quantum spin Hall insulators \cite{Fu09, Mi13} and chains of magnetic atoms,\cite{Cho11,Martin12,Nadj13,Klin13,Vazifeh13,Brau13} with recent experimental progress in both directions.\cite{Kne12, Du13, Yaz13} After the first proposals for braiding protocols in nanowire networks,\cite{Ali11,Hyart13,Sau10,Sau11,Hec12,Hal12} there 
is a need for a detailed analysis of the limitations which might hinder the braiding operation \cite{ Anton10, Che11,Sch13,Kar13} or cause decoherence of Majorana qubits.\cite{Chamon11, Cheng12, Budich12, Rainis12, Loss12b, Mazza,Konschelle13}

According to Anderson's theorem, electrostatic disorder has little influence in $s$-wave superconductors,\cite{And59} but in unconventional superconductors  it can induce subgap states at arbitrarily low energies.\cite{Olexei01} Indeed, electrostatic disorder is an unavoidable feature in experimental setups, and consequently much attention has been devoted to its impact on Majoranas.\cite{Bro00, Olexei01, Gru05, PotterLee, Brouwer11a, Brouwer11b,Stanescu11, Liu12, Bagrets12, Pikulin12, Lutchyn12, Sau12, Lob12, Takei13, Brouwer13, Wimmer13, DasSarma13, Fregoso} Importantly, Majorana end modes are found to be surprisingly robust against strong disorder despite the presence of localized low-energy bound states.\cite{Wimmer13}

\begin{figure}[t]
\includegraphics[width=0.96\columnwidth]{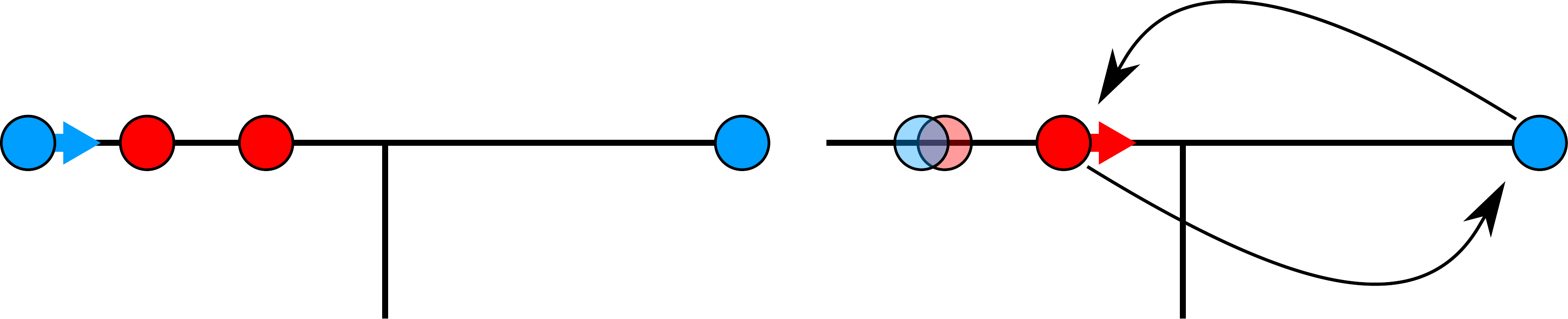}
\caption{(Color online) Detrimental effect of accidental Majorana modes (red) on a braiding manipulation: when a domain wall binding a computational Majorana mode (blue) approaches an accidental mode, these two Majoranas are fused. Quantum information is lost and the braiding protocol may proceed in a faulty manner, involving another accidental Majorana. 
}
\label{fig:braiding_failure}
\end{figure}

It is therefore important to investigate what happens to their non-Abelian statistics in the presence of disorder. To understand the potential problem, let us consider a disorder potential inducing two weakly coupled accidental Majorana modes, pinned to a particular location within the wire.\cite{Comment1} When a domain wall binding a computational Majorana moves towards an accidental one, the two modes couple strongly and  disappear into the continuum of states above the energy gap (see Fig.~\ref{fig:braiding_failure}). This fusion event leads to a loss of the information stored in the computational Majoranas.

Non-Abelian Majorana statistics can also be demonstrated using superconducting circuits \cite{Has11, Hec12, Hyart13} implementing an interaction-based braiding protocol.\cite{Bur13,Bon13} In these hybrid Majorana-transmon qubit devices, the braiding and readout protocols are realized by controlling Coulomb couplings between the Majoranas. In this Letter, we show that these protocols are efficiently realized even in the presence of disorder. We identify the dangerous physical processes  and show that the braiding errors are small if the couplings of the computational Majoranas to the accidental modes are much weaker than the maximum Coulomb coupling, leaving a large parameter space available for a braiding experiment.

The structure of the manuscript is as follows. We start in Section \ref{Sec-setup} by shortly reviewing the transmon circuit for the Coulomb-assisted braiding protocol, which was introduced in Ref.~\onlinecite{Hyart13}, and by presenting an effective model for the setup which captures the presence of disorder in the nanowries. In Sec. \ref{Sec-errors} we study numerically the time-evolution of the system during the flux-controlled protocol, and evaluate the effects of disorder on the braiding as well as on the initialization and measurement. To better streamline the presentation of results, we include some of the material as Appendices. We conclude with a few remarks in Sec. \ref{Sec-conclusion}.

\section{Braiding protocol in the presence of disorder}\label{Sec-setup}

To demonstrate non-Abelian statistics it is necessary to read out a topological qubit, described by the parity of two Majoranas $\Gamma_A$ and $\Gamma_B$, and to braid one of them, $\Gamma_B$, with another one, $\Gamma_C$. This task can be performed in a minimal fashion using a $\pi$-shaped nanowire network in a transmon circuit, following a flux-controlled braiding protocol.\cite{Hyart13} Although we consider Majoranas at the ends of nanowires, our results are applicable also to quantum spin Hall systems, where circuits can be constructed by using constrictions.\cite{Mi13}

\begin{figure}[!t]
\includegraphics[width=0.8\linewidth]{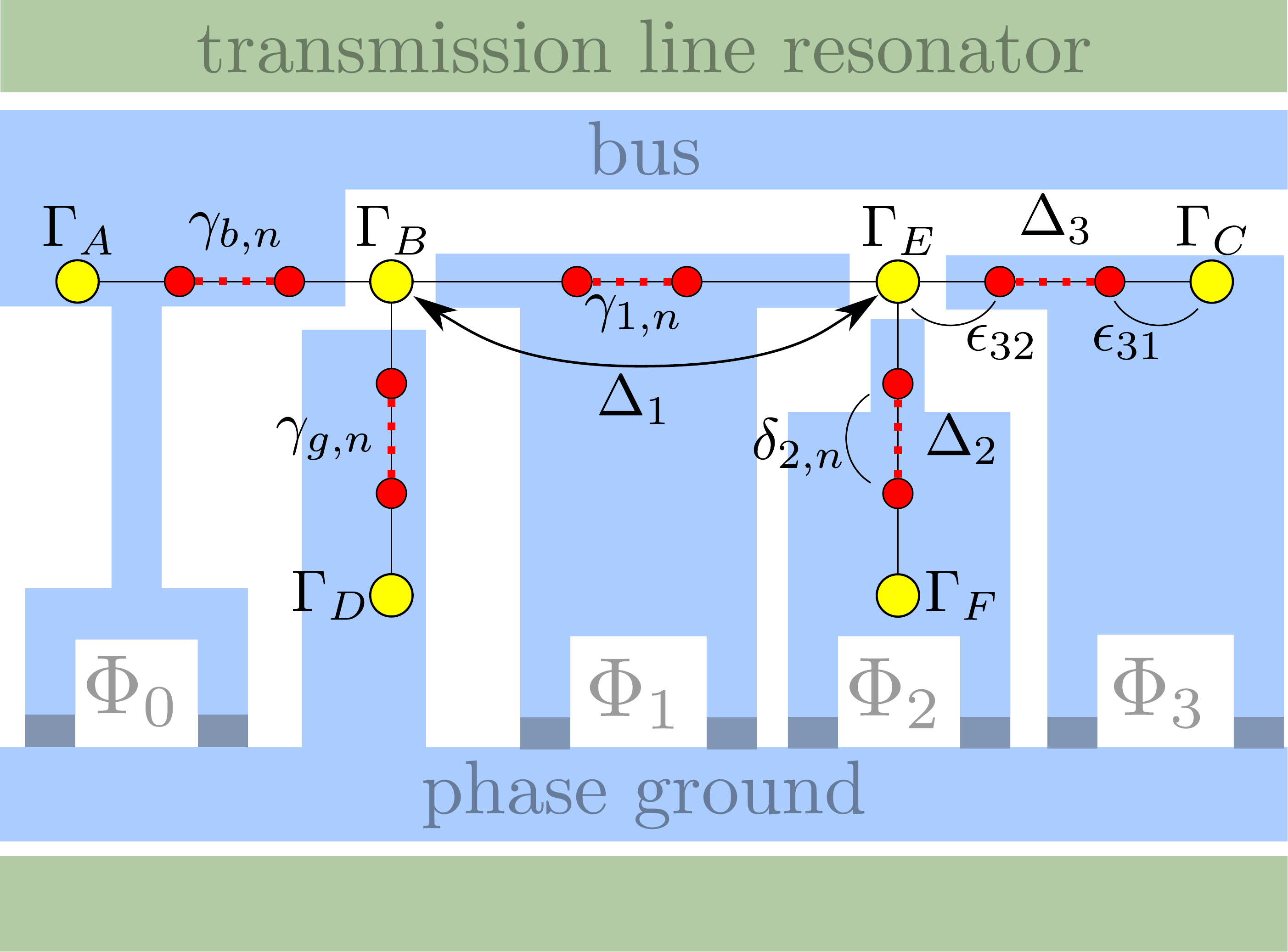}
\caption{Transmon circuit for  demonstration of  non-Abelian statistics.\cite{Hyart13} Two large superconducting islands (bus and ground) are used in the readout of the topological qubit and three smaller superconducting islands are needed for braiding. The nanowires form a $\pi$-shaped circuit  hosting six computational Majoranas, $\Gamma_A$, $\Gamma_B$,..., $\Gamma_F$. A strong disorder can induce accidental Majorana modes $\gamma_{k, n}$, where $k$ labels the island and $n$ the accidental Majorana mode within the island. These accidental modes are coupled to each other with couplings $\delta_{k, n}$, and the accidental Majoranas closest to the end of the wires are coupled to the corresponding end states with $\epsilon_{k1}$ and $\epsilon_{k2}$.}
\label{fig:layout}
\end{figure}

\begin{figure}[!t]
\includegraphics[width=\linewidth]{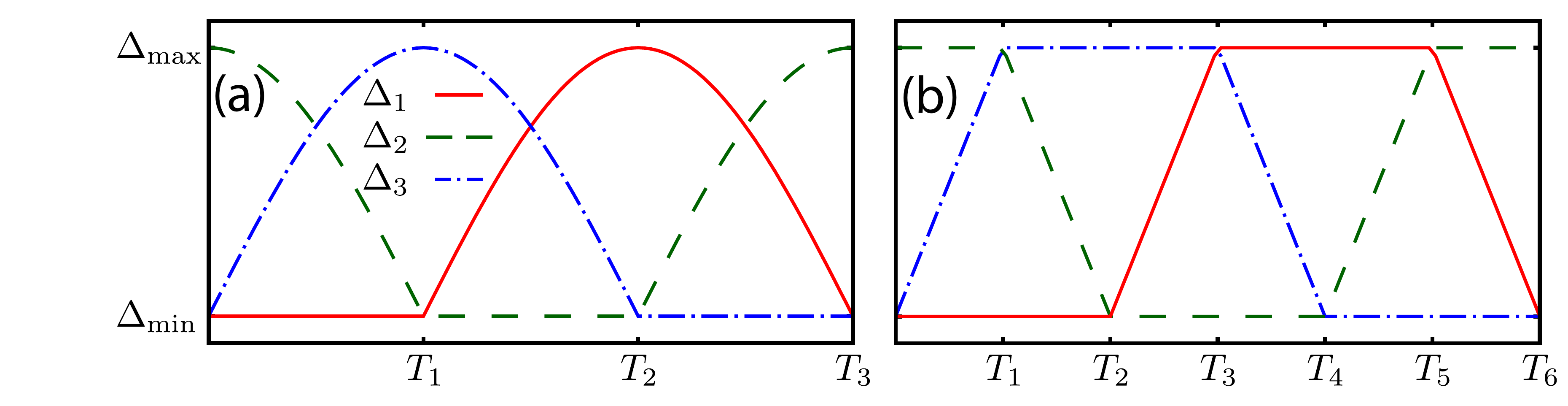}
\caption{Two possible paths of variations of Coulomb couplings resulting in braiding of Majorana zero modes $\Gamma_B$ and $\Gamma_C$. The braiding errors caused by the accidental modes depend on the braiding path (see Fig.~\ref{fig:errors}).}
\label{fig:paths}
\end{figure}

The circuit for braiding and readout is shown in Fig.~\ref{fig:layout}, and involves nanowires forming a $\pi$-shaped network hosting six computational Majoranas, $\Gamma_A$, $\Gamma_B$,..., $\Gamma_F$. The couplings between them can be controlled via the flux-dependence of the Josephson energy, $E_{J,k}(\Phi_k) = E_{J,k} (0) \cos(e \Phi_k /\hbar)$, of each superconducting  island, $k$.  The charging energies $E_{C, k}$ of the islands result in  Coulomb couplings $\Delta_k (\Phi_k)$ between the Majoranas, which, for $E_{J,k} (\Phi_k) \gg E_{C, k}$, have an exponential dependence $\Delta_k (\Phi_k) \propto \exp\big(-\sqrt{8E_{J,k}(\Phi_k)/E_{C,k}}\big)$,\cite{Has11, Hec12} that allows to turn them on ($\Delta_k=\Delta_{\rm max})$ and off  ($\Delta_k=\Delta_{\rm min})$ with fluxes.  A non-demolition readout of the topological qubit is possible, because the plasma frequency of the transmon formed by the bus and ground islands (see Fig.~\ref{fig:layout})  can be tuned close to the resonance 
frequency of the transmission line resonator.  Once the magnetic flux $\Phi_0$ is turned on, the coupling between photons and the transmon qubit renormalizes the resonance frequency of the cavity, so that it is conditioned on the fermion parity of $\Gamma_A$ and $\Gamma_B$.\cite{Has11, Hyart13} On the other hand, the Majorana modes $\Gamma_B$ and $\Gamma_C$ can be braided with the help of ancillas $\Gamma_E$ and $\Gamma_F$, by varying the Coulomb couplings $\Delta_k$ along a specific type of closed path \cite{Hec12} (see Fig.~\ref{fig:paths}). The corresponding operation on the topological qubit  is $\mathcal{U}=\exp(i s \pi \sigma_x/4)$,\cite{Iva01} where $s$ describes the braiding chirality. 

As we already pointed out, strong disorder induces accidental low-energy bound states in unconventional superconductors. 
These states can be described using Majorana operators $\gamma_{k, n}$, where $k$ labels the island and $n$ the accidental Majorana modes within it. We assume that neighboring Majoranas interact with random couplings. In particular, the accidental Majoranas closest to the end of each wire are coupled to the corresponding $\Gamma$ end modes with couplings $\epsilon_{k1}$ and $\epsilon_{k2}$ (see Fig.~\ref{fig:layout}). Unlike in the clean case, the Coulomb interaction involves the total fermion parity of each island, so braiding should be performed by controlling many-body interactions between Majoranas, instead of the simple pairwise ones considered in Refs.~\onlinecite{Hec12, Hyart13}. Similarly, the measurement is now sensitive to the total fermion parity of the bus island.

During the braiding procedure we set $\Phi_0 = 0$ so that the charging energy of the bus island can be neglected. The low-energy  Hamiltonian is
\begin{eqnarray}
H_{\rm br}&=&H_C+H_\delta+H_\epsilon  \\
H_C&=&  i \Delta_1\Gamma_B \Pi_1 \Gamma_E + i \Delta_2\Gamma_E \Pi_2 \Gamma_F + i \Delta_3 \Gamma_E \Pi_3 \Gamma_C,  \\
H_\delta&=& i \sum_{k,n} \delta_{k, n}\, \gamma_{k,n}\gamma_{k,n+1},  \\
H_\epsilon &=&  i \epsilon_{b1} \Gamma_A \gamma_{b,1}   + i \epsilon_{g1} \Gamma_B \gamma_{g,1} 
  + i \epsilon_{11} \Gamma_B \gamma_{1,1}   + i \epsilon_{21} \Gamma_E \gamma_{2,1} \nonumber \\
&&  + i \epsilon_{31} \Gamma_E \gamma_{3,1} +  i \epsilon_{b2} \gamma_{b,N_b } \Gamma_B  + i \epsilon_{g2} \gamma_{g,N_g} \Gamma_D  \nonumber \\ &&  + i \epsilon_{12} \gamma_{1,N_1 } \Gamma_E
+ i \epsilon_{22} \gamma_{2,N_2} \Gamma_F
+ i \epsilon_{32} \gamma_{3,N_3 } \Gamma_C,
\end{eqnarray}
where $H_C$ describes the Coulomb couplings between the Majoranas, and $H_\delta$, $H_\epsilon$ describe the tunnel couplings of the accidental Majoranas to each other, and to the computational ones, respectively. We have denoted the total parity of the accidental Majoranas in island $k$ with
$\Pi_k=\e^{-i\pi N_k/4}\prod_{n=1}^{N_k} \gamma_{k,n}$.

If $H_\epsilon=0$, then $[H_{\rm br}, \Pi_k]=0$,
which means that the computational and accidental Majoranas form two decoupled quantum systems. In a  sector of the eigenstates of $\Pi_k$ with eigenvalues $p_k$,  $H_C(\{p_k\})= i p_1 \Delta_1\Gamma_B \Gamma_E + i p_2 \Delta_2\Gamma_E  \Gamma_F + i p_3 \Delta_3 \Gamma_E  \Gamma_C$, which was considered in Refs.~\onlinecite{Hec12, Hyart13}. The Hilbert space is divided into ground and excited state manifolds, separated by an energy $2E_0$, where $E_0=\sqrt{\Delta_1^2+\Delta_2^2+\Delta_3^2}\ge \Delta_{\rm max}$. Because the braiding is performed adiabatically with respect to $\Delta_{\rm max}$, the transitions between these manifolds can be neglected and the time-evolution operator within each parity sector is
\begin{equation}
\mathcal{U}_0(\{p_k\}, p_{\rm anc})=e^{i s(\{p_k\}, p_{\rm anc})\pi \sigma_x/4} \prod_{i}   \mathcal{U}_{\rm int, i}(\{p_k\}, p_{\rm anc}), \label{Ueps0}
\end{equation}
where $\mathcal{U}_{\rm int, i}(\{p_k\},  p_{\rm anc})$ describes the internal time-evolution of the accidental Majoranas in island $i$, $s(\{p_k\}, p_{\rm anc})$ denote chiralities of the braiding in different sectors of the Hilbert space, and $p_\textrm{anc}$ is the parity of the ancillas $\Gamma_E$ and $\Gamma_F$.

We now assume that the measurement projects the system to an eigenstate of total parity on the bus island $\mathcal{P}= - i \Gamma_A \Pi_b \Gamma_B$.  (The requirements for a successful measurement are analyzed below.) 
The protocol for demonstrating non-Abelian Majorana statistics consists of a measurement $\mathcal{P}$ followed by $n$ braiding cycles, after which the parity is measured again. The probability of observing a parity flip after $n$ consecutive braidings, $p_{\rm flip} (n)$, is dictated by the Majorana statistics. For clean wires the sequence of probabilities is $p_{\rm flip} = 1/2,1, 1/2,0$ for $n = 1,2,3,4$, and it repeats itself periodically for larger values of $n$.\cite{Hyart13} Given Eq.~\eqref{Ueps0}, the sequence is independent on the accidental Majoranas and the initial state of the ancillas as long as $H_\epsilon=0$.  Thus, the only limitations in this case are quasiparticle poisoning and inelastic relaxation processes.\cite{Commentchirality}

\section{Analysis of the braiding protocol errors.}\label{Sec-errors}

\subsection{Effects of disorder on the braiding cycle}

The interaction $H_\epsilon$ between computational and accidental Majoranas may lead to fermion parity exchanges, giving rise to braiding errors. We assume that these coupling constants satisfy $\epsilon_{k1}, \epsilon_{k2} \ll \Delta_{\rm max}$, which allows to choose the braiding speed so that $\epsilon_{k1}, \epsilon_{k2} \ll \Delta_0 \ll \Delta_{\rm max}$, where the energy scale $\Delta_0=\hbar/T_0$ is determined by the duration $T_0$ of one segment of the braiding cycle in Fig.~\ref{fig:paths}. Thus, we can calculate the unperturbed time-evolution operator $U_0(t)$  in each parity sector using the adiabatic approximation and consider the effect of $H_{\epsilon}$ perturbatively. The total time-evolution operator for one braiding cycle can be written as
\begin{equation}
\mathcal{U}=\mathcal{U}_0+\sum_{k} \big[\frac{\epsilon_{k1}}{\Delta_0} \delta \mathcal{U}_{k1} +\frac{\epsilon_{k2}}{\Delta_0} \delta \mathcal{U}_{k2} \big], \label{errors}
\end{equation}
where $\mathcal{U}_0$ is the unperturbed time-evolution, which in different parity sectors is described by Eq.~\eqref{Ueps0}, and  $\delta \mathcal{U}_{k1,2}$ are corrections which can in principle be computed for an arbitrary disordered wire. These corrections couple the different parity sectors and can result in braiding errors.

Next, we analyze in detail the case where each nanowire contains a single pair of accidental Majorana modes, which are coupled to each other by $\delta$. This allows to identify the fundamental mechanisms of errors, which are present also in nanowires with many accidental Majorana modes.

\begin{figure}[t]
\includegraphics[width=0.5\columnwidth]{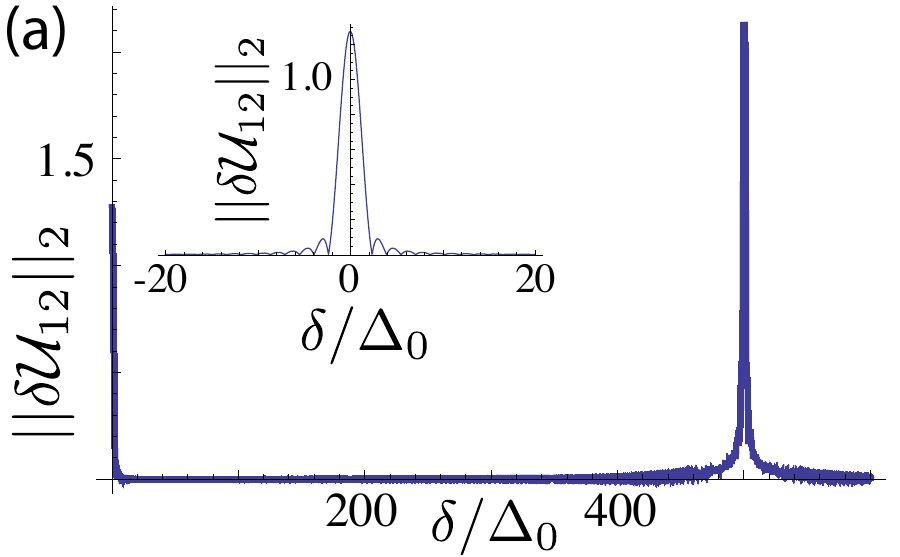}\includegraphics[width=0.5\columnwidth]{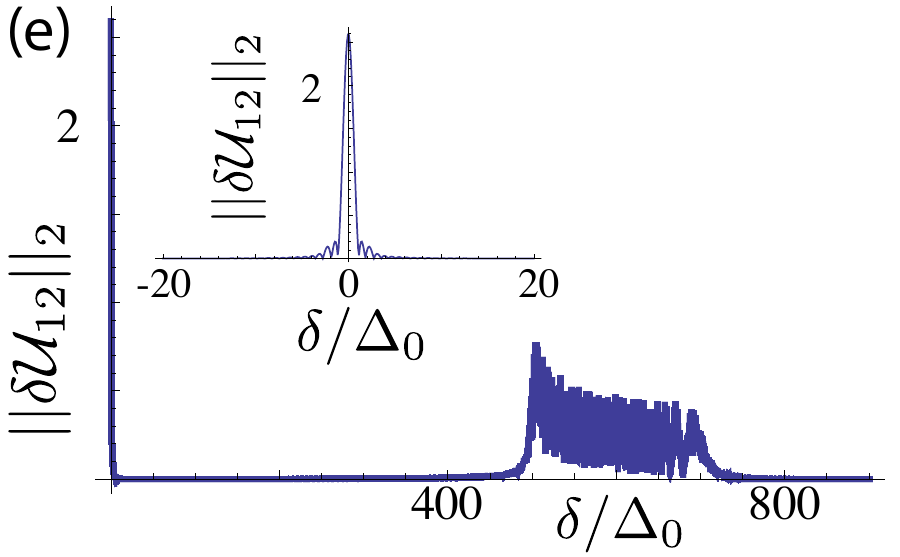}
\includegraphics[width=0.5\columnwidth]{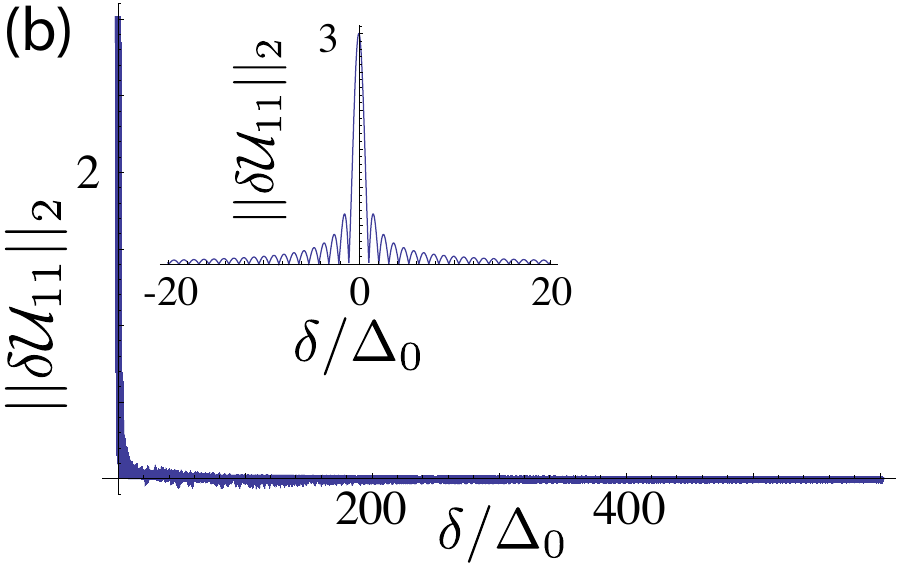}\includegraphics[width=0.5\columnwidth]{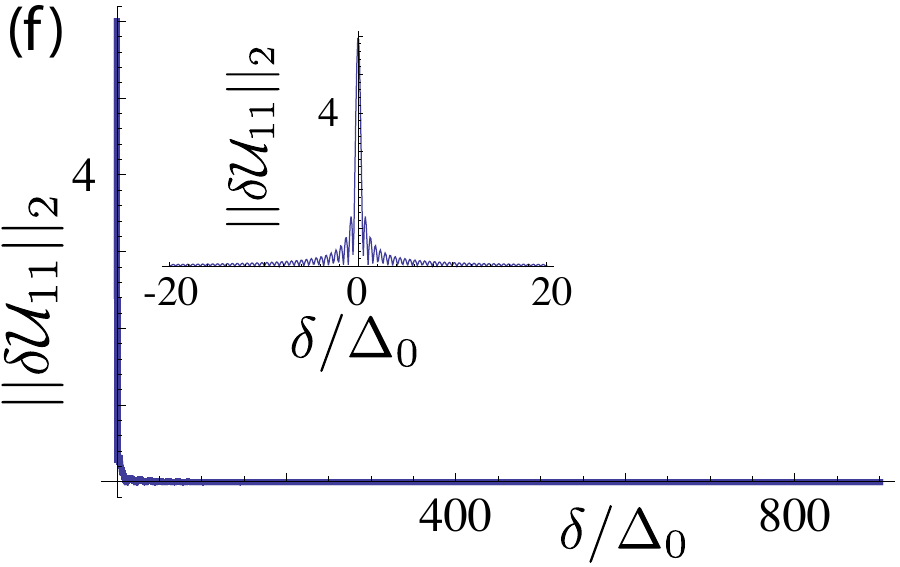}
\includegraphics[width=0.5\columnwidth]{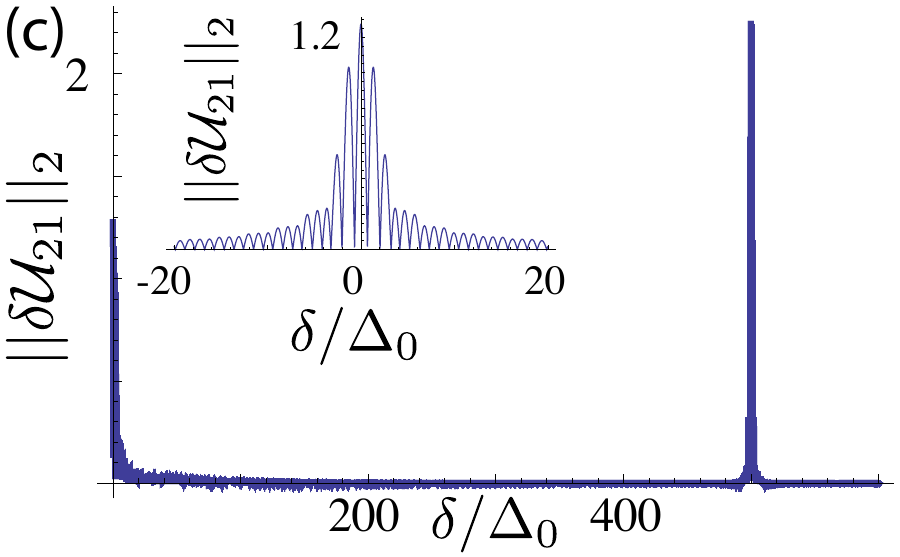}\includegraphics[width=0.5\columnwidth]{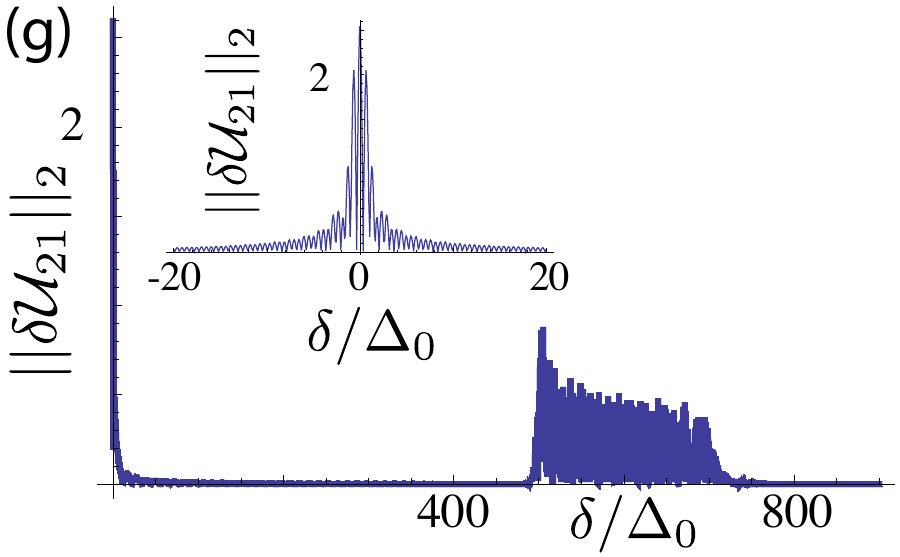}
\includegraphics[width=0.5\columnwidth]{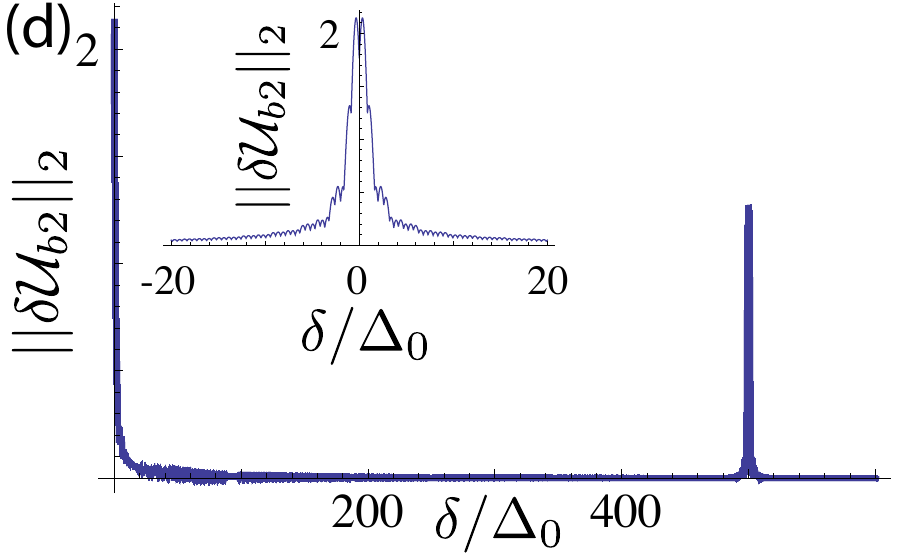}\includegraphics[width=0.5\columnwidth]{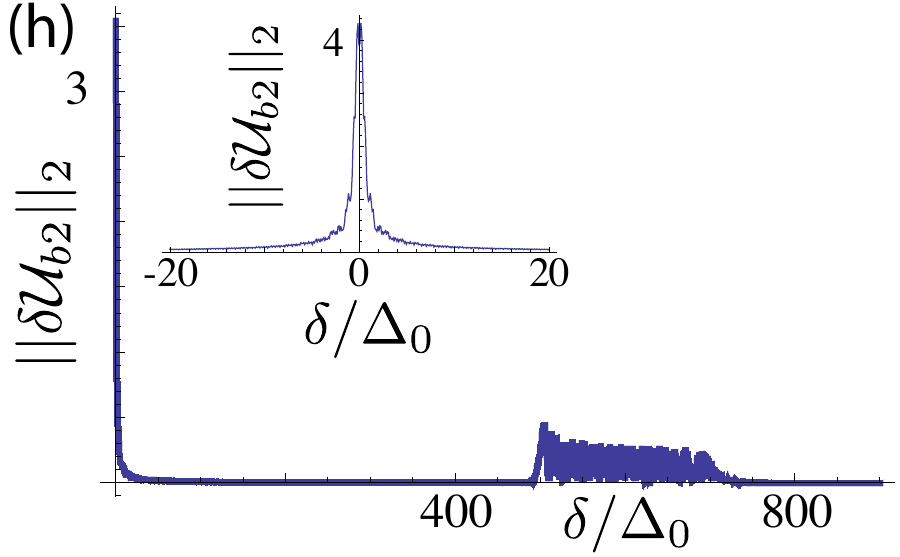}
\caption{Errors occurring during the braiding cycle can be estimated by $\epsilon_{ki}||\delta \mathcal{U}_{ki}||_2 /\Delta_0$  [Eq.~\eqref{errors}], with four different types of corrections $||\delta \mathcal{U}_{ki}||_2$, which are plotted as a function of $\delta$. These corrections, related to the two adiabatic cycles of Fig.~\ref{fig:paths}(a),(b), are shown in figures (a)-(d) and (e)-(h), respectively. The insets show magnifications of the peaks  around $\delta=0$. Away from the peaks the errors are efficiently suppressed. In all figures $\Delta_{\rm max}=500 \Delta_0$.}
\label{fig:errors}
\end{figure}

We first note that the couplings $\epsilon_{b1}$ and $\epsilon_{g2}$ have no effect on the braiding protocol within the lowest order perturbation theory. 
We characterize the errors caused by other couplings by calculating the matrix norms $||\delta \mathcal{U}_{ki}||_2$,\cite{comment_norm} which depend on $\delta$ and act as effective prefactors of ${\epsilon_{ki}}/{\Delta_0}$  in Eq.~\eqref{errors}. Based on symmetry arguments, we find that $||\delta \mathcal{U}_{b2}||_2=||\delta \mathcal{U}_{g1}||_2$, $||\delta \mathcal{U}_{11}||_2=||\delta \mathcal{U}_{22}||_2=||\delta \mathcal{U}_{32}||_2$ and $||\delta \mathcal{U}_{12}||_2=||\delta \mathcal{U}_{31}||_2$ (see Appendix \ref{AppA}). 
This leaves four different cases, which are plotted in Fig.~\ref{fig:errors} (a)-(d) and (e)-(h) for the two paths of Fig.~\ref{fig:paths}(a),(b), respectively. The errors show peaks when accidental Majorana fermions are either uncoupled ($\delta \approx 0$) or the energy of their bound state is in resonance with the energy gap between the ground and excited state manifolds ($\delta \approx E_0$). The peak appearing close to $\delta=0$ is extremely narrow for both paths, but the resonance at $\delta \approx \Delta_{\rm max}$ is strongly path dependent. 
For the circular path, shown in Fig.~\ref{fig:paths}(a), $E_0$ is constant during the whole braiding cycle resulting in narrow resonance peak at $\delta \approx \Delta_{\rm max}$. On the other hand, for the path shown in Fig.~\ref{fig:paths}(b), $E_0$ varies between $[\Delta_{\rm max}, \sqrt{2} \Delta_{\rm max}]$ during the braiding cycle so that the resonance peak spreads over a wide range of $\delta$. In the case of circular path it is possible to obtain closed form analytic solutions for $\delta \mathcal{U}_{ki}$. 
Away from the peaks where $||\delta \mathcal{U}_{ki}||_2 \sim 1$, they vanish asymptotically as $\sim \textrm{Max}\big[\Delta_0/\delta, \Delta_0/\big|\Delta_{\rm max} \pm \delta\big|\big]$ or faster (see Appendix \ref{AppB}). 
We have verified the validity of the perturbation theory for $\epsilon_{ki}/\Delta_0<0.1$ by numerically calculating the full time-evolution operator. We also point out that the assumption that the couplings $\delta_{k, n}$ and $\epsilon_{ki}$ are time-independent is not essential. Our qualitative findings are valid also if these couplings are changing adiabatically in time due to the variations of the Coulomb couplings. 

With increasing disorder, more low-energy sub-gap states will appear in the energy spectrum. For an increased number of accidental bound states, the braiding errors as a function of $\Delta_{\rm max}$ will contain several peaks, appearing whenever an energy of the accidental Majoranas is in resonance with $E_0$. This means that it becomes more and more difficult to avoid errors by properly choosing $\Delta_{\rm max}$. At the same time, the accidental modes will appear closer to the ends of the wires, increasing the couplings $\epsilon_{ki}$, which control the heights of the peaks in the braiding errors. As this coupling becomes comparable to the maximum Coulomb coupling, $\epsilon_{ki} \sim\Delta_{\rm max}$, one can no longer choose a $\Delta_0$ such that the braiding process is adiabatic with respect to the Coulomb coupling and non-adiabatic with respect to $\epsilon_{ki}$. At this point, the non-Abelian statistics is not observable anymore. An interesting theoretical question is whether this breakdown of 
the non-Abelian statistics happens in conjunction with a disorder-induced topological transition to a trivial phase of the nanowire, or whether it precedes it. We note that, in our model, the braiding process can in principle be optimized by choosing the coupling $\Delta_{\rm max}$ in such a way that it is comparable to the topological gap, $\Delta_{\rm max} \sim E_{\rm gap}$. In this case, non-Abelian statistics becomes unobservable when $\epsilon_{ki} \sim E_{\rm gap}$, so that the critical disorder strength is comparable to the critical disorder strength inducing the topological phase transition. However, our model is strictly speaking a low-energy effective theory, which is only valid in the nontrivial phase, and therefore it cannot be used for a detailed quantitative description of the breakdown of the non-Abelian braiding statistics and the topological phase transition happening at large disorder. 

\subsection{Effect of disorder on initialization and readout}

Errors can arise not only during the braiding cycle, but also during the readout, performed through a measurement of the fermion parity $\mathcal{P}= - i \Gamma_A \Pi_b \Gamma_B$ of the bus island. The Hamiltonian describing the interaction of the transmon qubit and the cavity is:\cite{Hyart13}
\begin{eqnarray} \label{JC}
  H_{\rm ro}&=& \hbar \omega_0 a^\dag a + \hbar g \left(\tau_+ a + \tau_- a^\dag \right) + \tau_z \left( \frac{1}{2}\hbar \Omega_0 + \Delta_+\mathcal{P} \right) \nonumber \\ && \hspace{-0.6cm}+ \Delta_-\mathcal{P} + H_b\left( \epsilon_{bn}, \delta_{b,n,m}  \right)+  i \epsilon_{11} \Gamma_B \gamma_{11}+  i \delta \gamma_{11} \gamma_{12}.
\end{eqnarray}
The first line describes the photons with bare resonance frequency $\omega_0$ and the interaction with the transmon qubit with a coupling constant $g$. Here $\Omega_0$ is the transmon plasma frequency, Pauli matrices $\tau_{x,y,z}$ act on the transmon qubit and  $\tau_\pm=(\tau_x \pm i\tau_y)/2$.  The term proportional to $\mathcal{P}$ arises due to the Coulomb coupling,\cite{Hyart13} and the Hamiltonian $H_b$ defines the tunnel couplings of the Majoranas inside the bus island. The last two terms describe the coupling of the computational Majorana $\Gamma_B$ to an accidental pair of modes outside the bus island. We assume that the transmission line resonator is  operated in the dispersive regime, where $(n+1)g^2\ll \delta\omega^2$, with $n$ the number of photons in the cavity and $\delta\omega=\Omega_0-\omega_0$.

Without accidental Majoranas, the Hamiltonian \eqref{JC} produces a parity-dependent resonance frequency of the cavity
$\omega_{\rm eff}({\cal P})={}\omega_0+\tau_z\,g^2(\delta\omega+2{\cal P} \Delta_+/\hbar)^{-1}$, which allows to measure the topological qubit.\cite{Hyart13, Has11} As before, we consider perturbative corrections caused by the couplings between computational and accidental Majoranas. The term $H_b$ conserves the parity $\mathcal{P}$ and therefore it does not modify $\omega_{\rm eff}$ within the lowest order perturbation theory. 
The presence of the external coupling $\epsilon_{11}$ implies that the measurement eigenstates of the renormalized cavity frequency no longer have a definite parity $\mathcal{P}$, but can be written in a form $\psi=\sqrt{1-\varepsilon^2}\ket{\mathcal{P},\ldots} + \varepsilon\ket{-\mathcal{P},\ldots}$, where away from resonances the measurement error vanishes as $\varepsilon \sim \epsilon_{11}/\big(\Delta_+-\Delta_-- |\delta| \big)$. This scaling is in agreement with the expected parity flow to the accidental Majorana modes. Close to the resonances  $\Delta_+-\Delta_- \approx |\delta|$ the parity flow will be limited by the finite measurement time $t_M$ so that the errors are $\sim \epsilon_{11} t_M/\hbar$. Therefore,  the conditions for successful measurement coincide with the requirements for  small braiding errors.

\section{Summary}\label{Sec-conclusion}

We have shown that the Coulomb-assisted braiding protocol is realizable also in the presence of disorder-induced accidental bound states, and that the braiding errors are small if the coupling of the computational Majoranas to the accidental states is much weaker than the maximum Coulomb coupling. A few remarks are in order concerning the experimental relevance of our results. First, the requirement of weak coupling between the computational and accidental Majorana modes coincides with the definition of the topological phase in disordered systems, and therefore based on the findings in Ref.~\onlinecite{Wimmer13}, we expect that there is a large parameter space available for braiding the Majorana fermions.  Secondly, the low-energy states in the wires can in principle be characterized using spatially resolved scanning tunneling microscopy \cite{Yaz13} or by coupling to microwaves.\cite{Schmidt13,Muller13,Ginossar13,Cottet13}
Because braiding errors depend strongly on the energies of the accidental modes, they can be systematically decreased by controlling these energies with the help of Zeeman fields or gate voltages. Finally, we point out that our results are relevant also in the case of clean wires, because they allow to simplify the experimental setup by replacing the $\pi$ shaped network of Ref.~\onlinecite{Hyart13} with two spatially separated T-junctions. In this case, two additional Majorana quasiparticles are intentionally created, which influence the braiding the same way as the accidental Majoranas considered here. However, in clean wires the additional Majoranas are automatically weakly coupled to the computational ones if the wires are sufficiently long, leading to negligible braiding errors.

\subsection*{Acknowledgements}

We have benefited from discussions with L. Kouwenhoven A. R. Akhmerov,  M. Wimmer and C. W. J. Beenakker. This work was supported by the Dutch Science Foundation NWO/FOM and by an ERC Advanced Investigator Grant.

\appendix
\section{Symmetry relations for the braiding errors}\label{AppA}

When the couplings between the accidental and the computational Majoranas is much smaller than the maximum Coulomb coupling, their effects can be treated independently. In the following we analyze each of the ten terms in $H_\epsilon$ and show that there are only four independent terms which contribute to errors during the braiding cycle.

Since the Coulomb Hamiltonian $H_C$  commutes with both $i\epsilon_{b1}\Gamma_A\gamma_{b,1}$, as well as $i\epsilon_{g2} \gamma_{g,N_g}\Gamma_D$, it is clear that these terms cannot cause errors during the braiding cycle. Their counterparts, $i\epsilon_{b2}\gamma_{b,N_b}\Gamma_B$ and $i\epsilon_{g1} \Gamma_B\gamma_{g,1}$ involve accidental Majoranas outiside the braiding T-junction and do cause errors, as shown in Fig.~\ref{fig:errors}. Furthermore, these errors are identical, $||\delta \mathcal{U}_{b2}||_2=||\delta \mathcal{U}_{g1}||_2$, because they are only related by a relabeling of the accidental Majorana indices.

Out of the six remaining terms, only three contribute in an independent fashion. To make this apperent, we will consider the case where there are only two accidental Majoranas in the braiding T-junction, which we label $\gamma_1$ and $\gamma_2$ for ease of notation. They may be placed in any of the three islands, and connected to any of the computational Majoranas. The six resulting Hamiltonians read:

\begin{widetext}
\begin{eqnarray}
  H_{11} & = & \Delta_1 \Gamma_B\gamma_1\gamma_2\Gamma_E +i\Delta_2\Gamma_E\Gamma_F+i\Delta_3\Gamma_E\Gamma_C + i\delta\gamma_1\gamma_2 + i\epsilon\Gamma_B\gamma_1 \\
  H_{12} & = & \Delta_1 \Gamma_B\gamma_1\gamma_2\Gamma_E +i\Delta_2\Gamma_E\Gamma_F+i\Delta_3\Gamma_E\Gamma_C + i\delta\gamma_1\gamma_2 + i\epsilon\gamma_2\Gamma_E \\
  H_{21} & = & i\Delta_1 \Gamma_B\Gamma_E +\Delta_2\Gamma_E\gamma_1\gamma_2\Gamma_F +i\Delta_3\Gamma_E\Gamma_C + i\delta\gamma_1\gamma_2 + i\epsilon\Gamma_E\gamma_1 \\
  H_{22} & = & i\Delta_1 \Gamma_B\Gamma_E +\Delta_2\Gamma_E\gamma_1\gamma_2\Gamma_F +i\Delta_3\Gamma_E\Gamma_C + i\delta\gamma_1\gamma_2 + i\epsilon\gamma_2\Gamma_F \\
  H_{31} & = & i\Delta_1 \Gamma_B\Gamma_E +i\Delta_2\Gamma_E\Gamma_F +\Delta_3\Gamma_E\gamma_1\gamma_2\Gamma_C + i\delta\gamma_1\gamma_2 + i\epsilon\Gamma_E\gamma_1 \\
  H_{32} & = & i\Delta_1 \Gamma_B\Gamma_E +i\Delta_2\Gamma_E\Gamma_F +\Delta_3\Gamma_E\gamma_1\gamma_2\Gamma_C + i\delta\gamma_1\gamma_2 + i\epsilon\gamma_2\Gamma_C.
\end{eqnarray}
\end{widetext}

Following Bravyi and Kitaev,\cite{Bravyi02} we write a representation of the six Majorana operators as:
\begin{eqnarray}\label{eq:gammasstart}
 \Gamma_B & = & \sigma_0 \otimes \sigma_0 \otimes \sigma_x \\
 \Gamma_C & = & \sigma_0 \otimes \sigma_0 \otimes \sigma_y \\
 \Gamma_E & = & \sigma_0 \otimes \sigma_x \otimes \sigma_z \\
 \Gamma_F & = & \sigma_0 \otimes \sigma_y \otimes \sigma_z \\
 \gamma_1 & = & \sigma_x \otimes \sigma_z \otimes \sigma_z \\
 \gamma_2 & = & \sigma_y \otimes \sigma_z \otimes \sigma_z,\label{eq:gammasend}
\end{eqnarray}
where $\sigma_i$ are the Pauli matrices and $\otimes$ denotes the Kronecker product.

The three Hamiltonians containing a coupling of an accidental Majorana to $\Gamma_B$, $\Gamma_F$, or $\Gamma_C$ are identical up to unitary transformations, and therefore lead to identical errors  $||\delta \mathcal{U}_{11}||_2=||\delta \mathcal{U}_{22}||_2=||\delta \mathcal{U}_{32}||_2$. The unitary transformations are
\begin{equation}
 H_{11} = U_{12}^{\phantom{\dag}} H_{22} U_{12}^\dag, \quad\quad
 H_{11} = U_{13}^{\phantom{\dag}} H_{32} U_{13}^\dag,
\end{equation}
with
\begin{eqnarray}
U^{\phantom{\dag}}_{12} = \begin{pmatrix}
           \sigma_z\otimes\sigma_z & 0 \\
           0 & \sigma_x \otimes \sigma_x
          \end{pmatrix}, \\
U^{\phantom{\dag}}_{13} = \begin{pmatrix}
           \sigma_z\otimes\sigma_z & 0 \\
           0 & \sigma_z \otimes \sigma_0
          \end{pmatrix}.
\end{eqnarray}
The Hamiltonians $H_{12}$ and $H_{31}$ can also be related by a unitary transformation, provided that one interchanges $\Delta_1$ and $\Delta_3$,
\begin{equation}
H_{12} = \widetilde{U}^{\phantom{\dag}}_{13} H_{31}(\Delta_1 \leftrightarrow \Delta_3) \widetilde{U}^\dag_{13}\,,
\end{equation}
where
\begin{equation}
\widetilde{U}^{\phantom{\dag}}_{13} =\frac{1}{\sqrt{2}}
\begin{pmatrix}
 i \sigma_0 \otimes (\sigma_x + \sigma_y) & 0 \\
 0 & \sigma_0 \otimes (\sigma_x + \sigma_y)
\end{pmatrix}\,.
\end{equation}
Since replacing $\Delta_1$ with $\Delta_3$ and vice versa amounts to performing the braiding cycle in a time-reversed order (see Fig.~\ref{fig:paths}), these two Hamiltonians produce identical errors $||\delta \mathcal{U}_{12}||_2=||\delta \mathcal{U}_{31}||_2$.

Such a tranformation also exists for $H_{12}$ and $H_{21}$, but involves replacing $\Delta_1 \leftrightarrow \Delta_2$, which changes the braiding path, and therefore leads to different errors, as shown in Fig.~\ref{fig:errors}.

\section{Analytical solutions for the braiding errors}\label{AppB}

In order to calculate the four independent corrections $||\delta \mathcal{U}_{ki}||_2$, we write the total time-evolution operator as $U(t) = U_0(t) \tilde{U}(t)$, where $U_0(t)$ is the time-evolution operator for $H_\epsilon=0$ and $\tilde{U}$ describes the lowest order correction 
caused by $H_\epsilon$. We assume that $\Delta_0 \ll \Delta_{\rm max}$, so that the unperturbed time-evolution operator $U_0(t)$ for the computational Majoranas in each parity sector can be calculated using the adiabatic approximation.  The lowest order correction $\tilde{U}$ can be found using the equation:
\begin{equation}
\tilde{U}(t)=1-\frac{i}{\hbar} \int _0^t d t_1 U_0^\dagger (t_1) H_\epsilon U_0(t_1). \label{correction}
\end{equation}
In this way we obtain that the total time-evolution operator for one braiding cycle is given by Eq.~\eqref{errors}, where $\mathcal{U}_0$ is the unperturbed time-evolution, which in different parity sectors is described by Eq.~\eqref{Ueps0}, and  $\delta \mathcal{U}_{k1}$ and $\delta \mathcal{U}_{k2}$ are corrections, which  can be solved by calculating the integral in Eq.~\eqref{correction}.

For the circular braiding path [Fig.~\ref{fig:paths}(a)] with one pair of accidental Majoranas in each island, the integral in Eq.~\eqref{correction} can be computed exactly, resulting in closed form analytic solutions for $\delta\mathcal{U}_{ki}$. Although the full expressions are not very insightful, they allow us to determine how the braiding error estimates, $\epsilon_{ki}||\delta \mathcal{U}_{ki}||_2 /\Delta_0$, vanish asymptotically far away from the resonant peaks in Fig.~\ref{fig:errors}. We obtain
\begin{widetext}
\begin{equation}
||\delta \mathcal{U}_{12}||_2 = {\rm Max}\bigg[\frac{\pi  |\cos(2 \delta/\Delta_0)|}{4 \delta^2/\Delta_0^2}, \frac{|\cos\big(3 (\delta \pm \Delta_{\rm max})/\Delta_0\big) \pm \sin\big(3 (\delta \pm \Delta_{\rm max})/\Delta_0\big)|}{\sqrt{2}|\delta \pm \Delta_{\rm max}|/\Delta_0} \bigg],
\end{equation}
\begin{equation}
||\delta \mathcal{U}_{11}||_2=\frac{|\sin(3 \delta/\Delta_0)|}{|\delta/\Delta_0|},
\end{equation}
\begin{equation}
 ||\delta \mathcal{U}_{21}||_2={\rm Max}\bigg[\frac{|\sin(3 \delta/\Delta_0)|}{|\delta/\Delta_0|}, \frac{\pi |\cos\big(3 (\delta \pm \Delta_{\rm max})/\Delta_0\big)|}{4(\delta \pm \Delta_{\rm max})^2/\Delta_0^2 }\bigg]
\end{equation}
and 
\begin{equation}
||\delta \mathcal{U}_{b2}||_2={\rm Max}\bigg[\frac{\sqrt{1 \pm \sin(6 \delta/\Delta_0)}}{\sqrt{2} |\delta/\Delta_0|},\frac{\pi |\cos(2 (\delta \pm \Delta_{\rm max})/\Delta_0)|}{4(\delta \pm \Delta_{\rm max})^2/\Delta_0^2}\bigg]. 
\end{equation}
\end{widetext}


\begin{thebibliography}{99}


\bibitem{Kit01} A. Yu. Kitaev, Phys. Usp. \textbf{44} (suppl.), 131 (2001).
\bibitem{Moo91} G. Moore and N. Read, Nucl. Phys. B \textbf{360}, 362 (1991).
\bibitem{Read2000} N. Read and D. Green, Phys. Rev. B \textbf{61}, 10267 (2000).
\bibitem{Iva01} D. A. Ivanov, Phys. Rev. Lett. \textbf{86}, 268 (2001).
\bibitem{Ali11} J. Alicea, Y. Oreg, G. Refael, F. von Oppen, and M. P. A. Fisher, Nature Phys. \textbf{7}, 412 (2011).
\bibitem{Ste04} A. Stern, F. von Oppen, and E. Mariani, Phys. Rev. B {\bf 70}, 205338 (2004).
\bibitem{Nay08} C. Nayak, S. H. Simon, A. Stern, M. Freedman, and S. Das Sarma, Rev. Mod. Phys. \textbf{80}, 1083 (2008).
\bibitem{Hyart13} T. Hyart, B. van Heck, I. C. Fulga, M. Burrello, A. R. Akhmerov, C. W. J. Beenakker, Phys. Rev. B {\bf 88}, 035121 (2013).
\bibitem{Lut10} R. M. Lutchyn, J. D. Sau, and S. Das Sarma, Phys. Rev. Lett. \textbf{105}, 077001 (2010).
\bibitem{Oreg10} Y. Oreg, G. Refael, and F. von Oppen, Phys. Rev. Lett. \textbf{105}, 177002 (2010).
\bibitem{Mou12} V. Mourik, K. Zuo, S. M. Frolov, S. R. Plissard, E. P. A. M. Bakkers, and L. P. Kouwenhoven, Science \textbf{336}, 1003 (2012).
\bibitem{Das12} A. Das, Y. Ronen, Y. Most, Y. Oreg, M. Heiblum, and H. Shtrikman, Nature Phys. \textbf{8}, 887 (2012).
\bibitem{Fu09} L. Fu and C. L. Kane, Phys. Rev. B {\bf 79}, 161408(R) (2009).
\bibitem{Mi13} S. Mi, D. I. Pikulin, M. Wimmer, and C. W. J. Beenakker, Phys. Rev. B {\bf 87}, 241405(R) (2013).
\bibitem{Cho11} T. P. Choy, J. M. Edge, A. R. Akhmerov, and C. W. J. Beenakker, Phys. Rev. B {\bf 84}, 195442 (2011)
\bibitem{Martin12}  I. Martin and A. F. Morpurgo, Phys. Rev. B {\bf 85}, 144505 (2012).
\bibitem{Nadj13} S. Nadj-Perge, I. K. Drozdov, B. A. Bernevig, and A. Yazdani, arXiv:1303.6393.
\bibitem{Klin13} J. Klinovaja, P. Stano, A. Yazdani and D. Loss, arXiv:1307.1442.
\bibitem{Vazifeh13} M. M. Vazifeh and M. Franz, arXiv:1307.2279.
\bibitem{Brau13} B. Braunecker and P. Simon, arXiv:1307.243.
\bibitem{Kne12} I. Knez, R.-R. Du and G. Sullivan, Phys. Rev. Lett. {\bf 109}, 186603 (2012).
\bibitem{Du13} L. Du, I. Knez, G. Sullivan, R.-R. Du, 	arXiv:1306.1925.
\bibitem{Yaz13} A. Yazdani \textit{et al.}, unpublished.
\bibitem{Sau10} J. D. Sau, S. Tewari, and S. Das Sarma, Phys. Rev. A \textbf{82}, 052322 (2010).
\bibitem{Sau11} J. D. Sau, D. J. Clarke, and S. Tewari, Phys. Rev. B \textbf{84}, 094505 (2011).
\bibitem{Hec12} B. van Heck, A. R. Akhmerov, F. Hassler, M. Burrello, and C. W. J. Beenakker, New J. Phys. \textbf{14}, 035019 (2012).
\bibitem{Hal12} B. I. Halperin, Y. Oreg, A. Stern, G. Refael, J. Alicea and F. von Oppen, Phys. Rev. B \textbf{85}, 144501 (2012).
\bibitem{Anton10} A. R. Akhmerov, Phys. Rev. B  \textbf{82}, 020509(R) (2010).
\bibitem{Che11} M. Cheng, V. Galitski, and S. Das Sarma, Phys. Rev. B {\bf 84}, 104529 (2011).
\bibitem{Sch13} M. S. Scheurer and A. Shnirman, arXiv:1305.4923 (2013).
\bibitem{Kar13} T. Karzig, G. Refael, and F. von Oppen, arXiv:1305.3626 (2013).
\bibitem{Chamon11} G. Goldstein and C. Chamon, Phys. Rev. B \textbf{84}, 205109 (2011).
\bibitem{Cheng12} M. Cheng, R. M. Lutchyn and S. Das Sarma,  Phys. Rev. B {\bf 85}, 165124 (2012).
\bibitem{Budich12} J. C. Budich, S. Walter and B. Trauzettel, Phys. Rev. B \textbf{85}, 121405(R) (2012).
\bibitem{Rainis12} D. Rainis and D. Loss, Phys. Rev. B \textbf{85}, 174533 (2012).
\bibitem{Loss12b} M. J. Schmidt, D. Rainis and D. Loss, Phys. Rev. B \textbf{86}, 085414 (2012)
\bibitem{Konschelle13} F. Konschelle and F. Hassler, arXiv:1306.2519.
\bibitem{Mazza}  L. Mazza, M. Rizzi, M. D. Lukin, and J. I. Cirac, arXiv:1212.4778.
\bibitem{And59}
P. W. Anderson, J. Phys. Chem. Solids {\bf 11}, 26 (1959).

\bibitem{Olexei01}
O. Motrunich, K. Damle, D. A. Huse, Phys. Rev. B {\bf 63},
224204 (2001).

\bibitem{Bro00}
P. W. Brouwer, A. Furusaki, I. A. Gruzberg, and C. Mudry, Phys. Rev. Lett. {\bf 85}, 1064 (2000).

\bibitem{Gru05}
I. A. Gruzberg, N. Read, and S. Vishveshwara, Phys. Rev. B {\bf 71}, 245124 (2005).

\bibitem{PotterLee}
A.C. Potter and P.A. Lee, Phys. Rev. B {\bf 83}, 184520 (2011).
\bibitem{Brouwer11a}
P. W. Brouwer, M. Duckheim, A. Romito, F. von Oppen, Phys. Rev. B {\bf 84},
144526 (2011).
\bibitem{Brouwer11b} P. W. Brouwer, M. Duckheim, A. Romito, and F. von Oppen, Phys. Rev. Lett. {\bf 107}, 196804 (2011).
\bibitem{Stanescu11}
T. D. Stanescu, R. M. Lutchyn, S. Das Sarma, Phys. Rev.
B {\bf 84}, 144522 (2011).
\bibitem{Liu12}
J. Liu, A. C. Potter, K.T. Law, P. A. Lee, Phys. Rev. Lett. {\bf 109}, 267002 (2012).
\bibitem{Bagrets12}
D. Bagrets and A. Altland, Phys. Rev. Lett. {\bf 109}, 227005
(2012).
\bibitem{Pikulin12}
D. I. Pikulin, J. P. Dahlhaus, M. Wimmer, H. Schomerus,
C. W. J. Beenakker, New J. Phys. {\bf 14}, 125011 (2012).
\bibitem{Lutchyn12}
R. M. Lutchyn, T. D. Stanescu, S. Das Sarma, Phys. Rev.
B {\bf 85}, 140513(R) (2012).
\bibitem{Sau12}
J. D. Sau, and E. Demler, arXiv:1204.2537.

\bibitem{Lob12}
A. M. Lobos, R. M. Lutchyn and S. Das Sarma, Phys. Rev. Lett. {\bf 109}, 146403 (2012).

\bibitem{Takei13}
S. Takei, B. M. Fregoso, H-Y Hui, A. M. Lobos, S. Das
Sarma, Phys. Rev. Lett. {\bf 110}, 186803 (2013).
\bibitem{Brouwer13}
M.-T. Rieder, P. W. Brouwer, I. Adagideli, arXiv:1302.2071.
\bibitem{Wimmer13}
I. Adagideli, M. Wimmer, A. Teker,  arXiv:1302.2612.
\bibitem{DasSarma13}
J. D. Sau and S. Das Sarma, arXiv:1305.0554.
\bibitem{Fregoso}
B. M. Fregoso, A. M. Lobos, S. Das Sarma, arXiv:1307.3505.

\bibitem{Comment1}
If we consider for example a sufficiently strong long-range correlated disorder, there will be accidental domain walls within the wire, giving rise to spatially well-separated Majorana modes with an exponentially weak coupling. Alternatively, we may consider a strong impurity within the wire, which pins a pair of zero energy Majoranas as shown in Ref.~\onlinecite{Sau12}.
\bibitem{Has11} F. Hassler, A. R. Akhmerov, and C. W. J. Beenakker, New J. Phys. \textbf{13}, 095004 (2011).
\bibitem{Bur13} M. Burrello, B. van Heck and A. R. Akhmerov, Phys. Rev. A \textbf{87}, 022343 (2013).
\bibitem{Bon13} P. Bonderson, Phys. Rev. B \textbf{87}, 035113 (2013).
\bibitem{Commentchirality} We note that while the accidental Majoranas do not influence the sequence  of probabilities $p_\textrm{flip}$, they affect the chirality of the braiding, which can be important in more advanced quantum manipulations.
\bibitem{comment_norm} The matrix norm $||\mathcal{U}||_2$ is defined as the largest singular value of $\mathcal{U}$.
\bibitem{Schmidt13} T. L. Schmidt, A. Nunnenkamp and C. Bruder, Phys. Rev. Lett. {\bf 110}, 107006 (2013).
\bibitem{Muller13} C. Muller, J. Bourassa, and A. Blais, arXiv:1306.1539.
\bibitem{Ginossar13} E. Ginossar and E. Grosfeld, arXiv:1307.1159.
\bibitem{Cottet13} A. Cottet, T. Kontos, and B. Douot, arXiv:1307.4185.
\bibitem{Bravyi02} S. Bravyi and A. Yu. Kitaev, Ann. Phys. (NY) 298, 210 (2002).


\end{thebibliography}
\end{document}